\documentstyle[11pt,paspconf,epsf]{article}

\begin{document}

\title{Resolving the Stellar Populations in a $z=4.04$ Lensed Galaxy}
\author{Andrew Bunker, Leonidas Moustakas, Marc Davis,\\
Brenda Frye, Tom Broadhurst \& Hyron Spinrad}
\affil{Department of Astronomy, University of California
at Berkeley,\\
601 Campbell Hall, Berkeley CA~94720 USA\\
{\tt email: bunker@bigz.Berkeley.EDU}}

\begin{abstract}
We have recently obtained deep near-IR Keck imaging of a
newly-discovered $z=4.04$ galaxy (Frye \& Broadhurst 1998). This is
lensed by the rich foreground cluster Abell~2390 ($z\approx 0.23$) into
highly-magnified arcs $3-5''$ in length.  Our $H$- and $K'$-band
Keck/NIRC imaging allows us to map the Balmer\,$+\,4000$\,\AA\ break
amplitude. In combination with high-quality archival HST/WFPC\,2 data,
we can spatially resolve stellar populations along the arcs. The WFPC\,2
images clearly reveal several bright knots, which may correspond to
sites of active star formation. Indeed, in some spatial regions the
Keck/LRIS discovery spectra are consistent with OB-star spectral energy
distributions in the rest-ultraviolet. However, there are considerable
portions of the arcs which appear redder with no Ly-$\alpha$ emission,
consistent with being post-starburst regions.
\end{abstract}

\section{Introduction}
Although $z>3$ galaxies are being found in increasing numbers ({\em
e.g.}, C. Steidel; J. Lowenthal, this volume), their study has been
almost exclusively through their ultraviolet (UV) flux redshifted into
the optical.  As such, we merely know about the near-instantaneous star
formation rate --- the UV continuum is dominated by massive short-lived
OB stars, and hence is sensitive to star formation on the time-scale of
only $10^7$ years. From studies of local galaxies, morphologies in the
UV can be very different from the appearance in the optical ({\em e.g.},
Colley {\em et al.\ }1996, ApJLett 473, 63). To
comprehensively study the underlying stellar populations, one should
also utilize longer rest wavelengths, where the light from older and
less massive stars may dominate.

A useful spectral feature in the study of stellar populations is the
4000\,\AA\ break, which arises from metal-line blanketing in late-type
stars. Even if the stellar population is only moderately evolved, there
will still be the Balmer break.  At high-$z$, these rest-frame optical
breaks are redshifted into the near-IR (Figure~1).  Here we present
preliminary results from an infrared study of stellar populations in a
newly-discovered lensed galaxy at $z=4.04$. At this redshift, in most
cosmologies the age of the Universe is less than $1\,h^{-1}_{100}\,$Gyr.
Evidence for any semi-evolved stellar population would have significant
implications for galaxy evolution in the early Universe.

\begin{figure}[ht]
{\centering \leavevmode \epsfxsize=0.5\textwidth \epsfbox{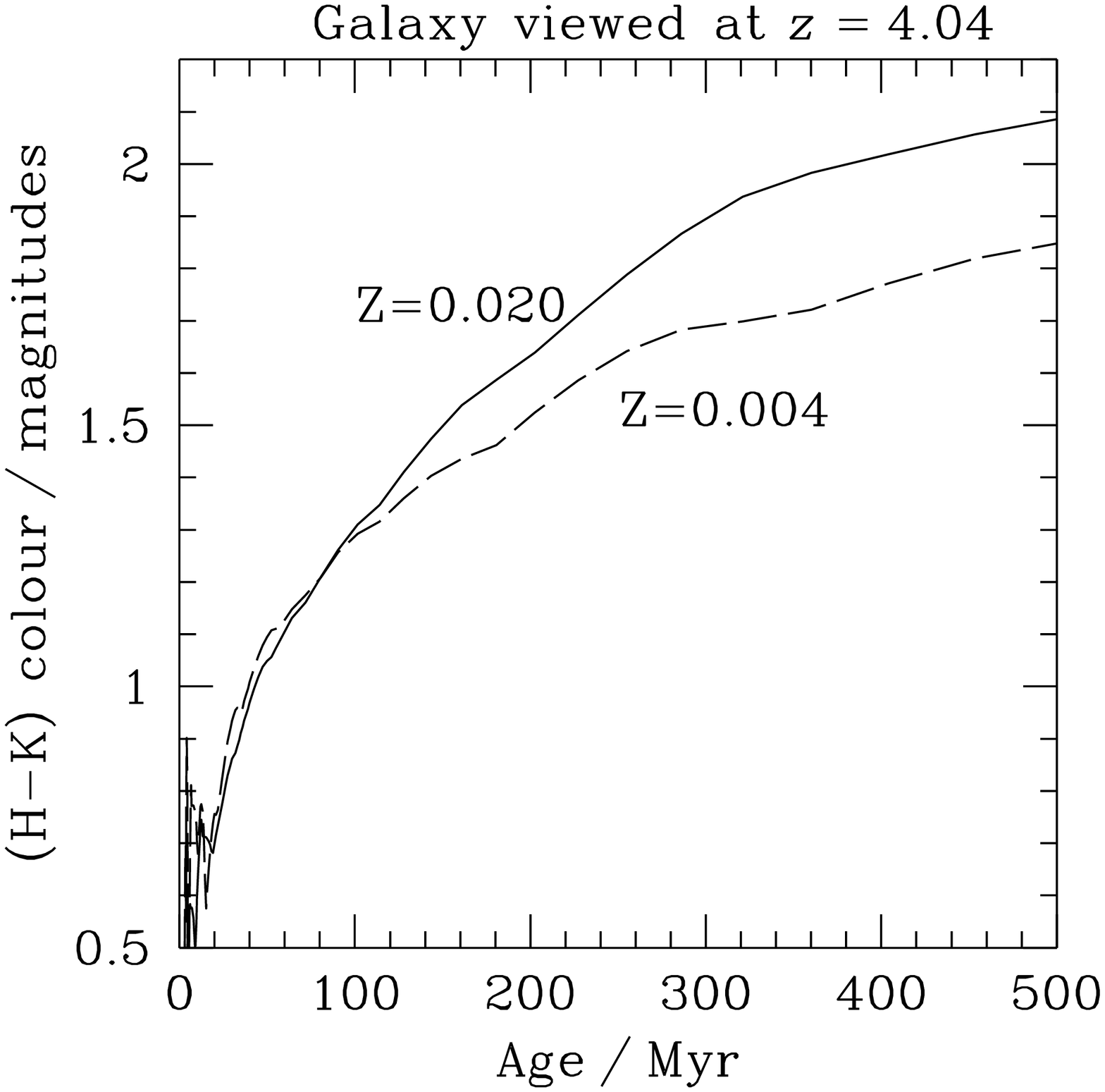}\hfil
   \epsfxsize=0.5\textwidth \epsfbox{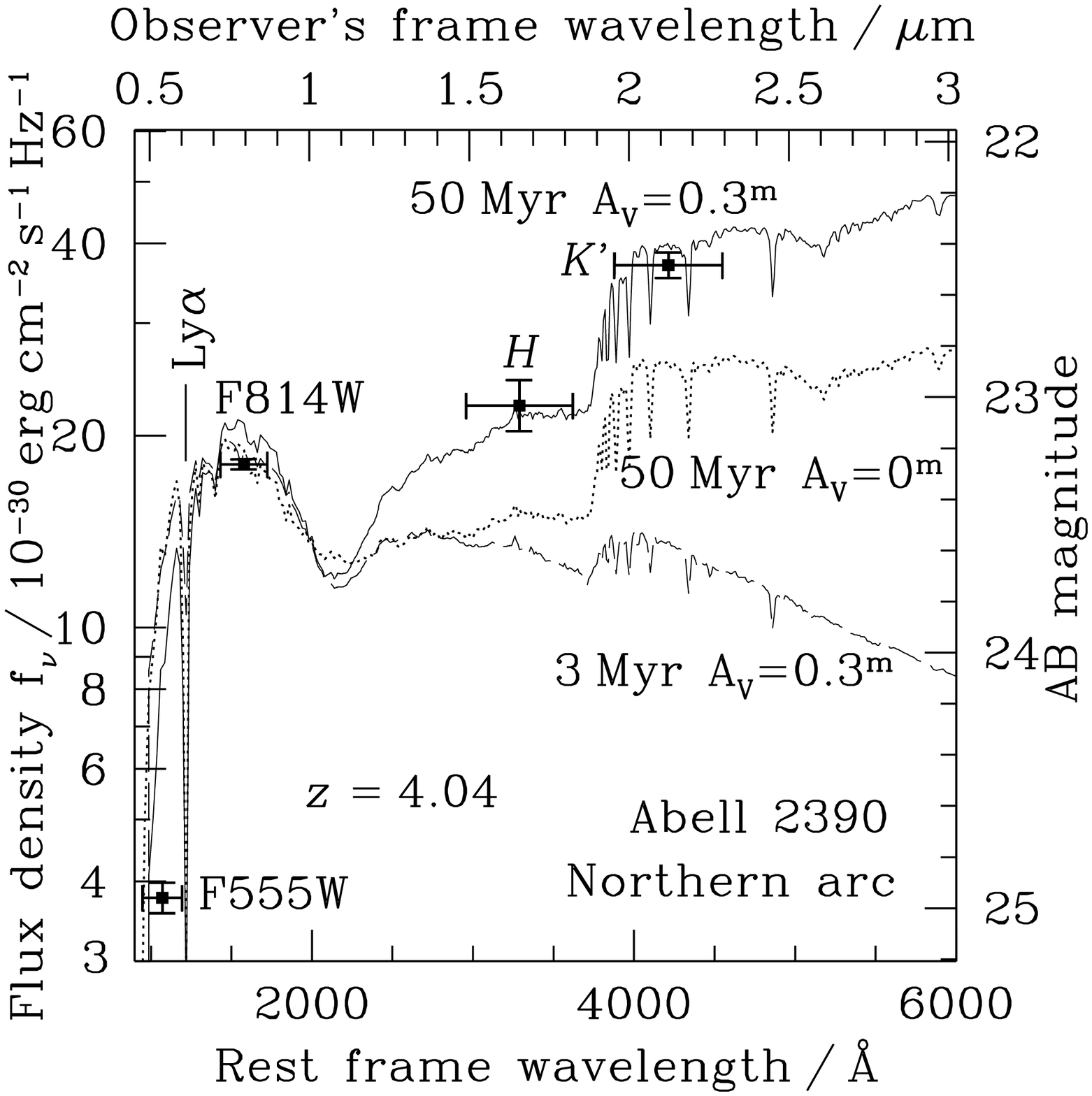}}

   \caption{{\bf Left:} The evolution of the $(H-K)$ colour of a galaxy
at $z=4.04$ as a function of the time elapsed since an instantaneous
burst of star formation. The solid curve is for Solar metallicity
($Z=0.020$), with the dashed line showing lower metallicity
($Z=0.004$). The models of Bruzual \& Charlot (1993, ApJ 405, 538) were
used for a Salpeter IMF with $0.1\,M_{\odot}<M^{*}<125\,M_{\odot}$.  For
this redshift, the $(H-K)$ colour is an excellent tracer of the time
elapsed since the end of star formation.
\newline {\bf Right:} The broad-band optical/near-IR
flux from the entire northern arc. Also plotted are
reddened instantaneous-burst stellar population models viewed at various
ages, arbitrarily normalized to the flux measured from the
HST/WFPC\,2 F814W image. The flux in F555W is severely
attenuated by the opacity of the intervening Ly-$\alpha$ forest.  The
colours are best reproduced by a stellar population $\sim 50$\,Myr old,
with {\em in situ} dust reddening of $A_{V}\sim 0.3^{m}$.  Note that at
$z=4.04$, the strong Balmer\,+\,4000\,\AA\ break due to the older stars
lies between the $H$- and $K'$-filters.}
\end{figure}

\begin{figure}[ht]
{\centering \leavevmode \epsfxsize=\textwidth \epsfbox{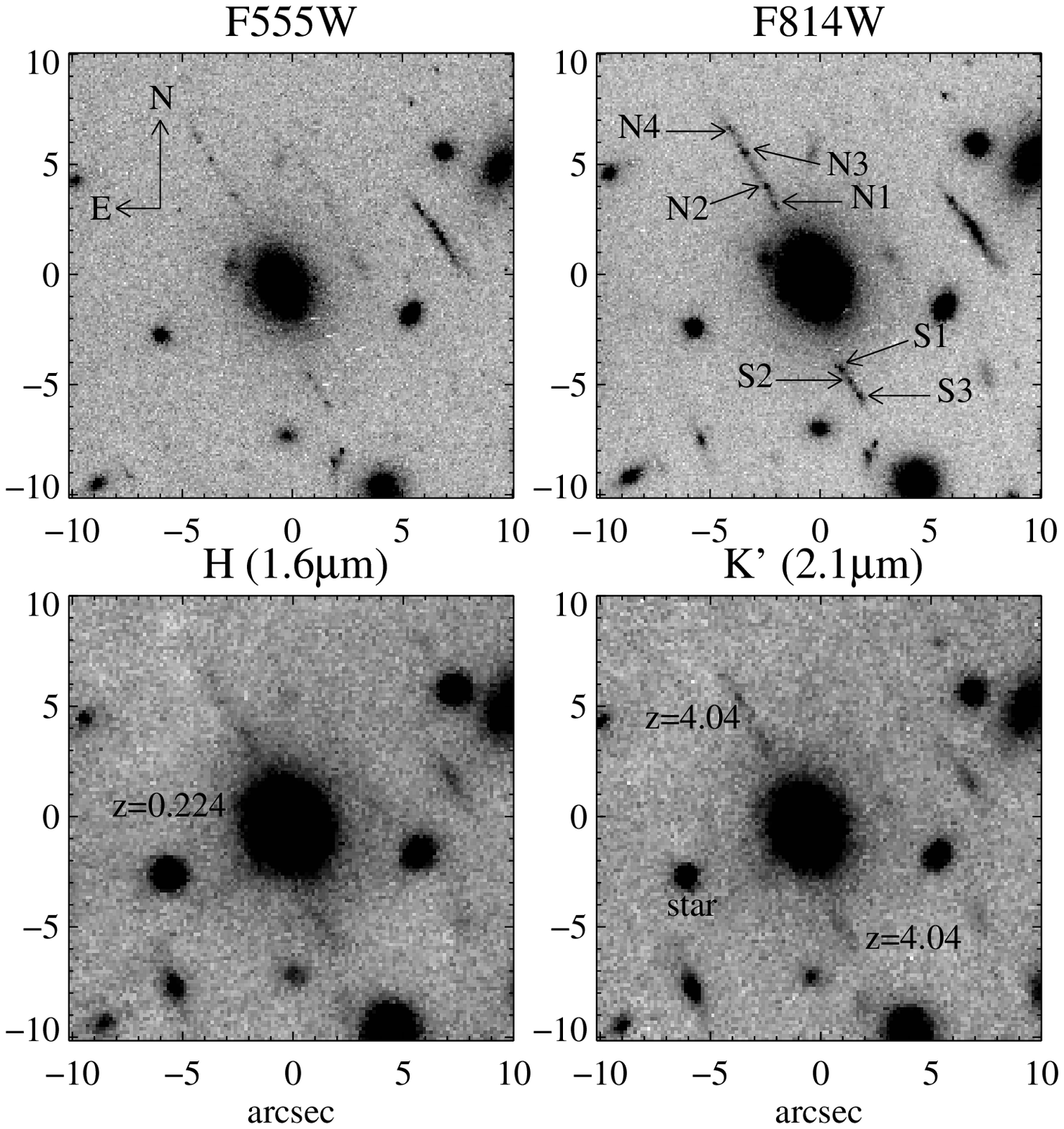}}
   \caption{The top panels show archival HST/WFPC\,2 imaging of the
   cluster Abell 2390, taken on 1994 December 10 U.T.\ (Fort {\em et
   al.\ }HST-GO\,5352). The elliptical in the centre of the field
   (RA=$21^{h}\,53^{m}\,33.52^{s}$,
   Dec.=$+17^{\circ}\,41^{'}\,57.4^{''}$, J2000) is at $z=0.22437$ (Yee
   {\em et al.\ }1996, ApJS 102, 289). The $z=4.04$ galaxy is the arclet
   at PA=$+23^{\circ}$ that is bisected by the elliptical, close to the
   Einstein ring $\sim 25''$ from the Abell~2390 cluster centre.  Top
   left is the HST $V$-band (F555W, 8400\,s) which encompasses
   Ly-$\alpha$, with the HST $I$-band (F814W, 10500\,s) top right. The
   knots which are bright in the rest-UV (and so are presumably sites of
   recent star formation) are indicated.  Our Keck/NIRC images were
   obtained on 1997 August 24 U.T.\ in good seeing ($0.4-0.5''$
   FWHM) and are shown lower left ($H$,
   2280\,s) and lower right ($K'$, 2880\,s).
The $K'$ filter was used in preference to the
   longer-wavelength $K$-band to avoid possible contamination by
   H$\beta$ line emission.}
\end{figure}

\section{A System of Lensed Arcs at $z\approx 4$}

Deep ground-based imaging and more recent HST studies of the cluster
Abell 2390 ($z\approx0.23$) have revealed many giant arcs, arising from
gravitational lensing of background galaxies.  In particular, a
$V\approx 20^{m}$ elliptical close to the critical curve of the cluster
bisects an extended pair of thin, near-straight arcs (Figure~2).  The
brighter northern arc ($I_{AB}=23.1^{m}$) is $\approx 5''$ long, with
the southern component ($I_{AB}=23.6^{m}$) $\approx 3''$ in length.  The
similarity in colours ($B-V>3^{m}$, objects 4\,\&\,5 in Pell\'{o} {\em
et al.\ }1991, ApJ 366, 405) and their positions relative to the caustic
suggest that these arcs are lensed images of the same source.  Imaging
with HST/WFPC\,2 reveals that the arcs are marginally resolved in the
radial direction ($\approx 0.3''$ FWHM).  Taking a na\"{\i}ve model of a
tangentially-stretched spherical source implies a magnification $>10$,
and more sophisticated lens geometries lead to a magnification of $\sim
25$ (Frye \& Broadhurst 1998, ApJLett {\em submitted}, astro-ph/9712111).  The
HST images also reveal the presence of two smaller counter-arcs.

 An arclet redshift survey by B\'{e}zecourt \& Soucail (1997, A\&A 317,
661) detected a single emission line at 6137\,\AA\ from the northern
component (their arclet\,\#20), and subsequent Keck/LRIS spectroscopy
(Frye \& Broadhurst 1998) confirmed this detection and identified the
line as Ly-$\alpha$ at $z=4.04$, on the basis of several rest-frame UV
interstellar absorption features. The southern arc and one of the
counter-arcs were also confirmed to be at $z=4.04$.  These arcs are
amongst the highest-redshift galaxies known, and are not pre-selected
for some AGN characteristic or extreme luminosity (they are magnified),
nor are they preferentially colour-selected for a young stellar population.
Thus, we may presuppose them to be lensed images of a `normal' galaxy,
and this affords us the opportunity to study stellar populations in the
early Universe, uncontaminated by any contribution from a central
engine.

\section{A Near-IR Study}

At $z\sim 4$, the rest-frame optical is redshifted to the near-infrared,
and the Balmer\,+\,4000\,\AA\ break is placed between the $H$-band (at
$1.65\,\mu$m) and the $K$-band (at $2.2\,\mu$m). In Figure~1 (left
panel) we plot the evolution of the $(H-K)$ colour with age for an
instantaneous starburst at $z=4.04$.
For this system, the $(H-K)$ colour is an excellent discriminator of the
stellar ages along the arcs.

We recently obtained deep $H$ \& $K'$ ($2.1\,\mu$m) imaging of the arcs
in good seeing with the Near Infrared Camera (NIRC) on the W.~M.\ Keck-I
Telescope (Figure~2).  The $K'$-band approximates to the rest-frame
$B$-band, and the integrated flux along the brighter northern arc
($K'=20.48^{m}$) suggests an intrinsic luminosity of
$L_{B}=0.3\,L^{*}_B \times (25/\mu)$ for a $q_{0}=0.5$ cosmology, where
$\mu$ is the magnification, and $0.8\,L^{*}_B \times (25/\mu)$ if
$q_{0}=0.1$.
Comparison with the WFPC\,2 images reveals colour gradients along both
arcs. Optical spectroscopy shows that only one of the four bright knots
of the northern arc has prominent Ly-$\alpha$ emission (labeled N4 in
Figure~2). The optical/near-IR colours of this are consistent with an OB
star, implying a region of active star formation.

However, the broad-band colours integrated along the entire northern arc
(Figure~1, right panel) show that the overall spectrum is best fit by a
more evolved ($\sim 50$\,Myr) population with extinction intrinsic to
the source of $A_{V}\sim 0.3^m$.  This implies that there are older
stars outside the actively star-forming H{\scriptsize~II} regions.  In
some areas, $(H-K)\sim 1.2^m$, indicating that $\sim100$\,Myr has
elapsed since the major episode of star formation (Figure~1, left).

Our Keck $H$- and $K'$-data allow us to `age-date' the stellar
populations in this spatially-resolved $z\approx 4$ galaxy by the
inferred amplitude of the Balmer\,+\,4000\,\AA\ break along the
transverse extent of the arcs. A more detailed treatment of the
near-IR/optical colours will be given in Bunker {\em et al.\ }(1998,
{\em in preparation}).

\acknowledgments

 We would like to thank the Organizing Committee. 
We are grateful to Daniel Stern for many useful discussions.
This work is based on
observations obtained at the W.M.~Keck Observatory, and with the
NASA/ESA Hubble Space Telescope, obtained from the data archive at STScI
which is operated by the Association of Universities for Research in
Astronomy, Inc., under the NASA contract NAS 5-26555. 

\end{document}